\documentclass[a4paper,11pt]{article}
\usepackage{pos}
\usepackage{aastex_hack}

\title{Exploring the population of Galactic very-high-energy $\gamma$-ray sources}
 
\author*[a]{Constantin Steppa}
\author[a]{Kathrin Egberts}
\affiliation[a]{Institut f\"ur Physik und Astronomie, Universit\"at Potsdam,\\
Karl-Liebknecht-Str. 24/25, Potsdam, Germany}

\forColl{CTA} 

\emailAdd{steppa@uni-potsdam.de}
\emailAdd{kathrin.egberts@uni-potsdam.de}

\renewcommand{\printHeadAuthors}{C. Steppa \& K. Egberts}

\abstract{At very high energies (VHE), the emission of $\gamma$ rays is dominated by discrete sources. Due to the limited resolution and sensitivity of current-generation instruments, only a small fraction of the total Galactic population of VHE $\gamma$-ray sources has been detected significantly. The larger part of the population can be expected to contribute as a diffuse signal alongside emission originating from propagating cosmic rays. Without quantifying the source population, it is not possible to disentangle these two components. Based on the H.E.S.S. Galactic plane survey, a numerical approach has been taken to develop a model of the population of Galactic VHE $\gamma$-ray sources, which is shown to account accurately for the observational bias. We present estimates of the absolute number of sources in the Galactic Plane and their contribution to the total VHE $\gamma$-ray emission for five different spatial source distributions. Prospects for CTA and its ability to constrain the model are discussed. Finally, first results of an extension of our modelling approach using machine learning to extract more information from the available data set are presented.}

\FullConference{37$^{\rm{th}}$ International Cosmic Ray Conference (ICRC 2021)\\
		July 12th -- 23rd, 2021\\
		Online -- Berlin, Germany}

\begin{document}
\maketitle

\section{Introduction}
In the last decades, we have continuously expanded our horizon for the very high-energy universe. Yet, with about 150 VHE $\gamma$-ray sources discovered, we have uncovered only a small fraction of the population of Galactic sources. This small sample is not representative of the entire population, as it has a strong inherent observational bias. The composition of this sample is the result of the complex interplay between global source distributions, such as source position, luminosity or physical extent, and instrumental selection effects. In this contribution we describe a model of the population of Galactic VHE $\gamma$-ray sources derived from the source sample of the H.E.S.S. Galactic plane survey (HGPS) catalogue \cite{Hess2018} using a population synthesis method to correctly account for the observational bias. The model was originally presented in \cite{Steppa2020}. For more information beyond the summary given here, please read the original publication.
\section{Model}
In our model, we treat each source as a generic $\gamma$-ray emitter, i.e. we do not model specific classes of sources. This approach is complementary to detailed source modelling and allows us to determine global properties that we need, for example, to estimate the contribution of unresolved sources to the measurement of large-scale diffuse emission. The model is composed of independent distribution functions for the position of the sources, their luminosity and their physical extent.
\\For the source position, we took five different spatial distribution functions with fixed parameters from the literature to investigate what implications this has for the other properties of the population. Next to two azimuthally symmetric distributions, three distributions following a spiral-arm structure were also chosen. Both the luminosity and the physical extent---represented here only by the radius of a spherically symmetric object---are described by simple power laws with fixed limits, so that the indices of these power laws are the only free parameters of the model.
\section{Parameter estimation}
The parameters of the model were reconstructed using population synthesis in combination with maximum likelihood estimation. First, the observation bias of the HGPS was estimated from the sensitivity map provided with the catalogue. In doing so, the description of the instrument's sensitivity was extended to account for its degradation with the angular extent of the sources. With this estimate, the expected number of detected sources can be calculated for any given global source distribution, so that the model parameters were optimised based on a likelihood determined by Poissonian count statistics. The accuracy of this procedure was validated with Monte Carlo simulations of synthetic populations. The final result of this optimisation is shown in Table~\ref{TAB:fit}.
\begin{table}[!tb]
\centering
\begin{tabular}{lcc}
\hline
Spatial distribution & $\alpha_{L}$ & $\alpha_{R}$
\\
\hline
mSNR & -1.70 & -1.19 \\
mPWN & -1.81 & -1.13 \\
mFE & -1.94 & -1.21 \\
mSp4 & -1.64 & -1.17 \\
mSp2B & -1.78 & -1.62 \\
\hline
\end{tabular}
\caption{Estimates of the indices of the luminosity and radius distribution functions, $\alpha_{L}$ \& $\alpha_{R}$, calculated based on different spatial distributions of the sources. Details are given in \cite{Steppa2020}.}
\label{TAB:fit}
\end{table}
\section{Model predictions}
To assess our model and parameter estimation, we compared distributions of observable quantities such as source flux or position on the sky of the HGPS sample with simulated sources of synthetic populations that fall within the sensitivity range of the HGPS. In general, good agreement is achieved between observation and model, with the exception of a spatial distribution model that assumes a two-arm spiral structure. Also, the observed asymmetry in the source distribution over Galactic latitude and the distribution over angular extent does not yet find satisfactory agreement in the model. Despite these discrepancies, the model allows us to gain insight into the Galactic population of VHE $\gamma$-ray sources. Depending on the spatial distribution model, the estimated total number of sources is between 800 and 7000. If the actual population is close to the upper limit of this range, this challenges the standard paradigm that supernova remnants and pulsar wind nebulae are the dominant source classes of VHE $\gamma$-ray emission.
\\The combined flux of unresolved sources in the region covered by the HGPS was estimated to be $13\,\%$-$32\,\%$ of the flux of the total source sample in this region. This magnitude is compatible with the measurement of large-scale diffuse emission by H.E.S.S. \cite{hess2014} and a clear indication that the contribution of unresolved sources is indeed not negligible. A template of unresolved sources that we can generate with our population synthesis approach will allow us to assess this contribution more precisely.
\\Next-generation instruments such as CTA will be able to resolve a significant fraction of the yet unresolved sources and help us improve our knowledge of the global population. With a targeted sensitivity for point sources of $2\,\mathrm{mCrab}$ in the longitude range $|l|<60^{\circ}$ and the latitude range $|b|<2^{\circ}$ \cite{cta2019} and an angular resolution of $0.05^{\circ}$ at $1\,\mathrm{TeV}$, the predicted number of detectable sources is in the range 200-600, which increases the known Galactic VHE $\gamma$-ray source sample by a factor of between $5$-$9$ (within this region).
\section{Outlook}
The resolution of the entire population of Galactic VHE $\gamma$-ray sources will be beyond our capabilities in the foreseeable future. Therefore, population models are an important tool to assess measurements of VHE $\gamma$-ray emission and the applied analysis techniques. In our continuous efforts to improve our population model, we have developed a new, more precise method for reconstructing the model parameters based on Deep Learning. This method reduces the reconstruction errors by about a factor of two. More details on this method will be given in a future publication. Also, in future iterations of the model, we will look more closely at the impact of source confusion on the obtained sample in the sensitivity range. Rough estimates suggest that $22\,\%$-$24\,\%$ of the sources in the HGPS are already affected by source confusion and similar values are expected for the CTA GPS. While the effect of source confusion on the sample of detected sources and their measured properties remains to be evaluated, adequately accounting for this effect in the modelling is the logical next step towards a more accurate description of the global population.
\acknowledgments
This work was conducted in the context of the CTA Consortium. We gratefully acknowledge financial support from the agencies and organisations listed here: \url{http://www.cta-observatory.org/consortium_acknowledgments}.

\bibliographystyle{JHEP}
\bibliography{proceeding}

\clearpage
\section*{Full Authors List: The Cherenkov Telescope Array Consortium}

H.~Abdalla\textsuperscript{1}, H.~Abe\textsuperscript{2},
S.~Abe\textsuperscript{2}, A.~Abusleme\textsuperscript{3},
F.~Acero\textsuperscript{4}, A.~Acharyya\textsuperscript{5}, V.~Acín
Portella\textsuperscript{6}, K.~Ackley\textsuperscript{7},
R.~Adam\textsuperscript{8}, C.~Adams\textsuperscript{9}
% ...

1 : Centre for Space Research, North-West University, Potchefstroom, 2520, South Africa

2 : Institute for Cosmic Ray Research, University of Tokyo, 5-1-5, Kashiwa-no-ha, Kashiwa, Chiba 277-8582, Japan

3 : Pontificia Universidad Católica de Chile, Av. Libertador Bernardo O'Higgins 340, Santiago, Chile

4 : AIM, CEA, CNRS, Université Paris-Saclay, Université Paris Diderot, Sorbonne Paris Cité, CEA Paris-Saclay, IRFU/DAp, Bat 709, Orme des Merisiers, 91191 Gif-sur-Yvette, France

5 : Centre for Advanced Instrumentation, Dept. of Physics, Durham University, South Road, Durham DH1 3LE, United Kingdom

6 : Port d'Informació Científica, Edifici D, Carrer de l'Albareda, 08193 Bellaterrra (Cerdanyola del Vallès), Spain

7 : School of Physics and Astronomy, Monash University, Melbourne, Victoria 3800, Australia

8 : Laboratoire Leprince-Ringuet, École Polytechnique (UMR 7638, CNRS/IN2P3, Institut Polytechnique de Paris), 91128 Palaiseau, France

9 : Department of Physics, Columbia University, 538 West 120th Street, New York, NY 10027, USA

% ...
%
%
%

\end{document}